\documentclass{article}

\usepackage{arxiv}

\usepackage[utf8]{inputenc} 
\usepackage[T1]{fontenc}    
\usepackage{hyperref}       
\usepackage{url}            
\usepackage{booktabs}       
\usepackage{amsfonts}       
\usepackage{nicefrac}       
\usepackage{microtype}      
\usepackage{lipsum}
\usepackage{graphicx}

\usepackage{color}
\usepackage{xspace}

\graphicspath{{./figures/}}

\newcommand{\showfig}[1]{#1}

\newcommand{\comment}[1]{}

\begin{document}

\title{Neuromodulation influences synchronization and\\ intrinsic read-out}

\author{Gabriele Scheler\\
Carl Correns Foundation for Mathematical Biology,\\
1030 Judson Dr., Mountain View, Ca 94040\\
\texttt{gscheler@gmail.com} \\
}

\maketitle

\begin{abstract}
\textbf{Background}: The roles of neuromodulation in a neural network, such as in a cortical microcolumn, are still incompletely understood. Neuromodulation influences neural processing by presynaptic and postsynaptic regulation of synaptic efficacy. Neuromodulation also affects ion channels and intrinsic excitability. 

\textbf{Methods}: Synaptic efficacy modulation is an effective way to rapidly alter network density and topology. We alter network topology and density to measure the effect on spike synchronization. We also operate with differently parameterized neuron models which alter the neuron's intrinsic excitability, i.e., activation function. 

\textbf{Results}: We find that (a) fast synaptic efficacy modulation influences the amount of correlated spiking in a network. Also, (b) synchronization in a network influences the read-out of intrinsic properties. Highly synchronous input drives neurons, such that differences in intrinsic properties disappear, while asynchronous input lets intrinsic properties determine output behavior. Thus, altering network topology can alter the balance between intrinsically vs. synaptically driven network activity.

\textbf{Conclusion}: We conclude that neuromodulation may allow a network to shift between a more synchronized transmission mode and a more asynchronous intrinsic read-out mode. This has significant implications for our understanding of the flexibility of cortical computations.
\end{abstract}

%

\section{Introduction}
In this paper we present a realistic network model, akin to a cortical
microcolumn 
\cite{Douglas2004,Douglas2007,Carlo2013,Kunze2017}, 
and
investigate its properties under the assumption of fast synaptic and
intrinsic modulation as evidenced by neuromodulation 
\cite{Scheler2004g}.
We hypothesize that rapid synaptic efficacy changes allow a network to operate with
%
%
different topologies, and that network topology is a decisive
factor towards creating and sustaining synchronized inputs vs.
producing asynchronous input. 

We have previously shown for a conductance-based neural model of
striatal medium spiny neurons that neuronal heterogeneity expressed by
the contribution of individual ion channels (such as delayed rectifier
potassium channels or GIRK channels) may still result in
uniform responses, if the neurons are driven with highly correlated
synaptic input. If the same neurons are driven by more asynchronous,
distributed synaptic input, the heterogeneity is manifest in the
response patterns, i.e. the spike rates and the timing of the spikes
(see \cite{Scheler2014}).
These results were achieved using conductance-based point
neurons \cite{Scheler2014}.
Here we use two-dimensional neural models 
\cite{Izhikevich2004a}
to further investigate the effect and determine its significance in 
the context of a cortical neural network.

Due to Hebbian learning 
\cite{Scheler2017,Koulakov2009}, 
under normal conditions synaptic weights
follow a lognormal distribution, which results in graphs with a heavy
tail degree distribution. 
Degree modification by rapid synaptic efficacy changes would not only allow 
 for alterations to the density, but also
the topology of the connecting graph. In this paper we examine the
hypothesis that such changes in network topology actually occur, driven
by neuromodulatory effects on presynaptic release or postsynaptic
response \cite{Varela2009,Ohshima2017,Kobayashi2009,Scheler2004g}.
We analyze this situation with two example graphs, and we
also perform further analysis to show that there is a
continuum of graphs which can be reached by rapid synaptic changes.

\section{Methods}
\subsection{Conductance-based neuron model and synaptic input}

The conductance-based neural model of a striatal medium spiny neuron is
described in detail in \cite{Scheler2014}. The membrane voltage $V_m$ is modeled using
the equation
\begin{equation}
\dot{V}_m = - {1\over C}[\mu_1(I_1) + \mu_2(I_2) + \ldots + \mu_n(I_n) - I_{syn}],
\end{equation}
where the $I_i$ are the currents, induced by the individual
ion channels. Variability of the neuron is modeled by modifications to
$\mu_i$. This model includes ion channels for Na (INa), K (IK), slow
A-type K channels (IAs), fast A-type K channels (IAf), inward
rectifying K channels (IKir), L-type calcium channels (ICaL), and the
leak current (I\textsubscript{leak}). The definition of all parameters
and the dynamics of the ion channels can be found in \cite{Scheler2014}.

For the experiments in this paper, we use only a single channel as an example for  the variability that can be induced
by neuromodulatory changes. We chose the slow A-type K channels as in \cite{Scheler2014}. The total
current contribution for this channel is $\mu_{\mbox{\footnotesize IAs}}$
where $\mu$ was
selected between 1.0 and 1.5, a variability by $\pm 25\%$.

In order to illustrate the variability in neuron behavior, we excited
the neuron model by input signals, resembling two kinds of synaptic
input: uncorrelated and correlated. These signals were generated by
superposition of excitatory and inhibitory spikes from individual
Poisson-distributed spike trains (50 excitatory and 10 inhibitory), and
biased Gaussian background noise. The details can be found in \cite{Scheler2014}.
The amount of pairwise correlation in these spike trains governs the
type of input signal. A high correlation factor was used in order to
generate sequences which have short periods (10--15ms) of high
activity.

\subsection{Heterogeneity in a two-dimensional model}

In order to do large-scale simulation we needed to employ a simple,
computationally tractable neuron model. We used a two-dimensional model
of a neural oscillator (cf.\ \cite{Izhikevich2004a}), and employed
an instantiation of the model with parameters fitted to the general
properties of cortical pyramidal neurons \cite{Izhikevich2004} as a generic model (g). 
The model consists of an equation for the membrane model $v$
(Equation~\ref{eq1}), fitted
to experimental values for cortical pyramidal neurons, and an equation
for a gating variable $u$ (Equation~\ref{eq2}).

\begin{equation}
\dot{v} = 0.04 v^2 + 5 v + 140 - u - I_{syn}
\label{eq1}
\end{equation}
\begin{equation}
\begin{array}{l}
\dot{u} = a(bv - u)\\
b = 0.2 \\
a = 0.02 \\
\end{array}
\label{eq2}
\end{equation}

When the neuron fires a spike (defined as $v(t) = 30 mV$), $v$ is
set back to a low membrane potential
$v := c; c = -65.8mV$
and the gating variable $u$ is increased by a fixed amount $d$ 
($u := u+d; d = 8$) (cf.~\cite{Izhikevich2004}). 
This formulation allows for a very simple
neuron model, which avoids the explicit modeling of the downslope of
the action potential, and rather resets the voltage. 
Time-dependence after a spike is modeled by the gating variable $u$.

Neuronal heterogeneity is achieved by systematic variation of
inactivation parameters. By varying $d$, we can vary the
inactivation dynamics of the model after a spike, by varying $a$
we vary the activation/inactivation dynamics for $u$.
In this
way, we can attempt to model neuronal variability in
activation/inactivation dynamics, which is sufficient to model
frequency-selectivity as a stored intrinsic property. The parameters
used in this paper for different neuron types are listed in Table~\ref{tab1}.

\begin{table}[h!]
\caption{\label{tab1}Parameters for different neuron types (cf.\cite{Izhikevich2004})}
\centering
\begin{tabular}{l|llll}
name & $a$ & $b$ & $c$ & $d$\\
\hline
generic &  	0.02 &	0.2 &	-65 &	8\\
type1 &	0.025&	0.2 &-65 &6	\\
type2 &	0.02 &0.2&-65 &9\\
type3 &	0.015&	0.2&	-65&	12\\
type4 &	0.015&	0.15&	-65&	14\\
type5&	0.022&0.3 &-65 &14\\
type6&		0.022&	0.3 &-65&	9.5\\
\end{tabular}
\end{table}

\subsection{Graph properties}

We created graphs of $N$ ($=210$) excitatory neurons,  and $K$ ($\approx 1900$)
excitatory connections.  For the excitatory neurons, we use randomly
connected graphs (N,K) with different width 
$\sigma^* = e^\sigma$ of the probability distribution. 
This corresponds to normal (Gaussian) to
lognormal graphs with different widths and length of the heavy tail. We
model inhibition by Poisson-distributed inhibitory synaptic input
directly onto excitatory neurons.

We use specific instantiations of these graphs (RG, LG1) for the
simulations.
Table~\ref{tab2} shows global graph characteristics for the Gaussian 
graph (RG), the lognormal graphs (LG1), and intermediate graphs LG2, LG3, and LG4.

\begin{table}[h!]
\caption{\label{tab2}Graph Properties for Model Networks (Excitatory Connections)}
\centering
\begin{tabular}{l|lllll}
Property & RG &	LG1& 	 LG2   &         LG3 &         LG4 \\
\hline
N no of nodes &  210   & 210        &       210   & 210      &  210 \\
K connections&  1880  &  1924      &        1895  &     1920&  2050 \\
d density& 0.042  &       0.043 &  0.043& 0.043 & 0.047 \\
indegree&  8.95[2...17]  &         9.5[0...26] &          9[0...24]&  9.14[0...30] & 9.75 [0...25] \\
outdegree& 8.95[4...18] &          9.5[0...56]&         9[0...58]&  9.14[0...53]&  9.75 [0...54] \\
cluster index&  0.04  &  0.068 &  0.064  &   0.062    &   0.08 \\
mean path length& 2.66    &            2.41  &  3.7 &     3.72  &  2.32 \\
width $\sigma^*$   & 1.44 &  2.89&    2.5   &   2.71 &2.71 \\
synchronization $s$ &      0.11  &                     0.32    &                   0.26  &       0.31  &      0.34 \\
\end{tabular}
\end{table}

The rewiring algorithm used to change the properties of a graph G is a 
greedy algorithm, which iteratively selects the node with the highest 
degree. 
One of its edges is then rewired to random nodes with lower degrees, 
decreasing $\sigma^*$. The algorithm terminates, when the value of 
$\sigma^*$ falls below a given threshold. 

\subsection{Definitions}

We define synchronization $s$ in a network by pairwise correlations: 
for each neuron $n_i$, we count, for each other neuron $n_j$, the number 
of spikes which occur within a window $W$ ($W = 10ms$) of $n_i$'s 
spike events, divided by the total number of spikes for $n_i$. 
More precisely, for each neuron, we bin all firing events into 5ms bins. 
We then count the number of spikes emitted by other neurons, 
which fire in a 10ms window around 
the (start) of the bin.
The synchronization $s$ is then the average over all neuron pairs 
in the network:

$$
s=\frac{1}{N(N-1)}\sum _{j}{\frac{1}{S_{j}}}\sum _{i}B_{\mathit{ij}},$$

where  $B_{\mathit{ij}}$ is the number of spikes that neurons \textit{i}
and \textit{j} have in common within a moving window of $W=10ms$ during
the entire measuring time. $S_{j}$ is the number of spikes of neuron
\textit{j} during the entire measuring time.

\subsection{Simulation Tools}
All simulations were performed with the software tool \href{https://gscheler.github.io/CNeuroSim/}{CNeuroSim}, which  is implemented in \href{https://uk.mathworks.com/products/matlab.html}{Matlab} (R2016b) and C, and available at 
%
%
\url{https://doi.org/10.5281/zenodo.1164096}.

\section{Results}

\subsection{
Conditional expression of intrinsic excitability in conductance-\-based
models}

We show how we can model gain as a stored intrinsic property, defined as
the average spike rate in response to constant input (constant input in A / average spike rate in Hz).
We used a full ion channel based model (the MSN model \cite{Scheler2014}), with
variation in the slow A-type potassium channel (IAs). This ion channel was used as an example for the conductance-based model \cite{Scheler2014}. Neuromodulation often affects just one type of ion channel, and a total variations of 30-50\% in ion channel efficacy have been typically found.

\begin{figure*}[!h]
\showfig{
\centering
{\bf A}
\includegraphics[width=0.40\textwidth]{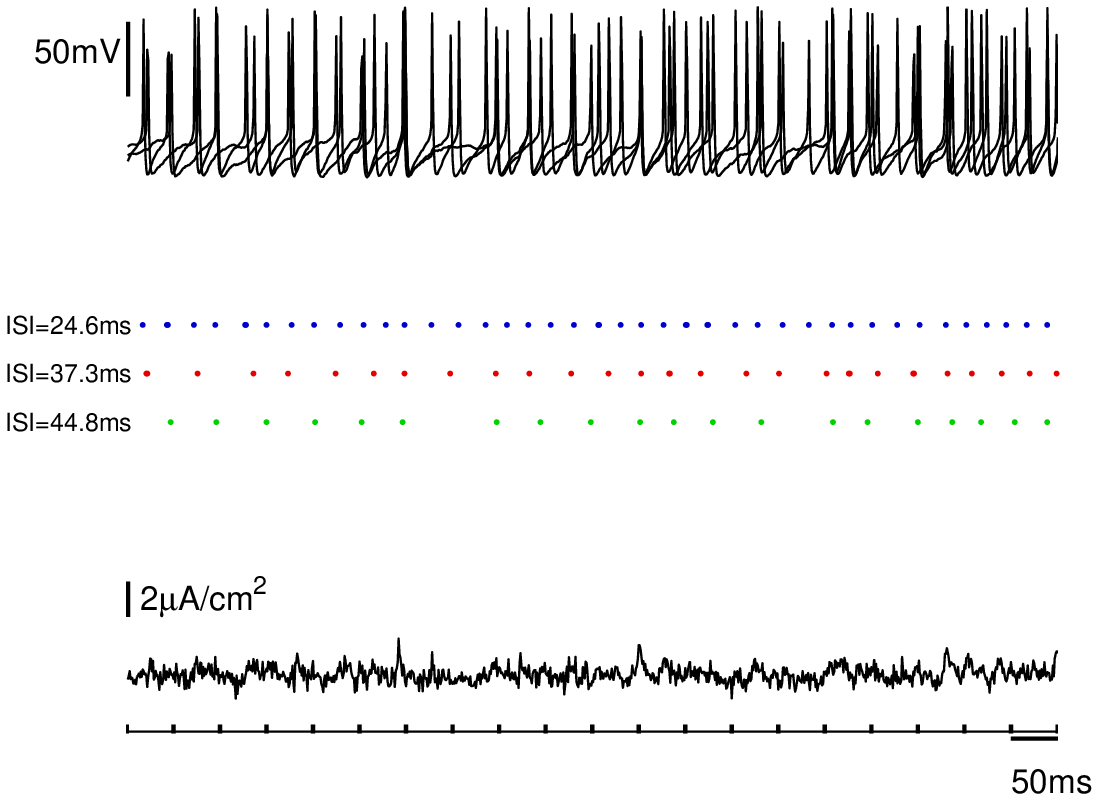}
\hspace*{1cm}
{\bf B}
\includegraphics[width=0.40\textwidth]{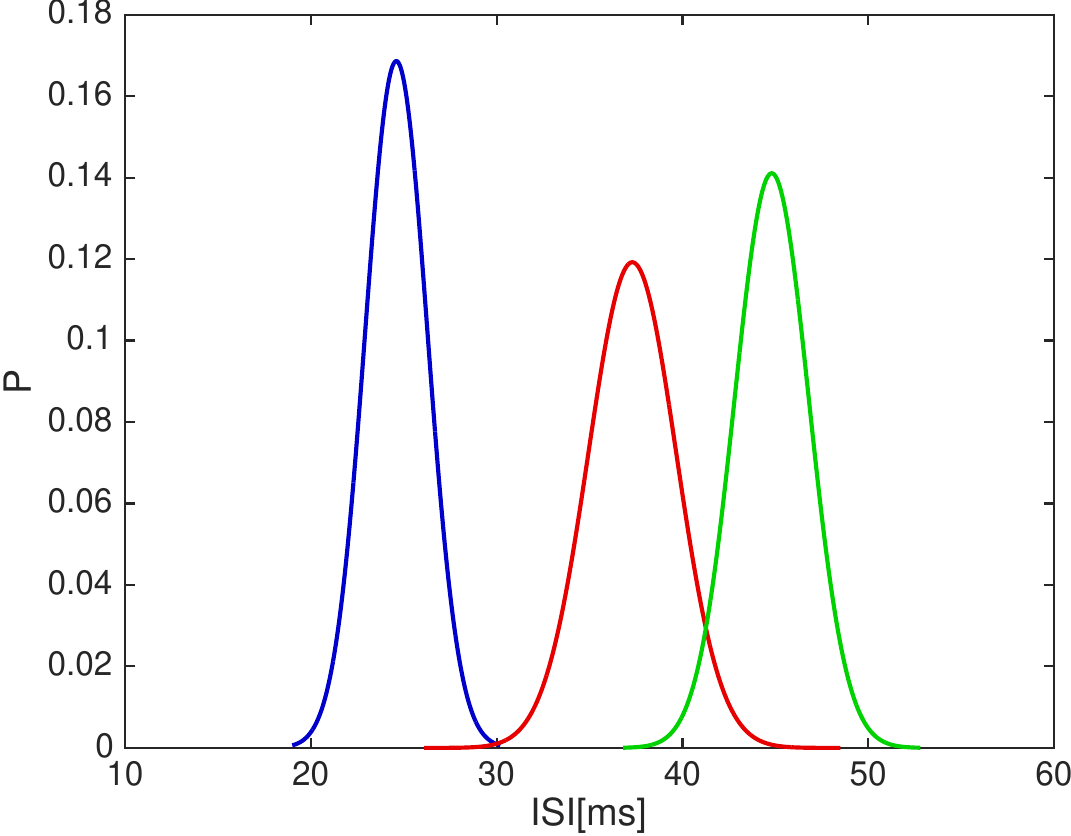}
}
\caption{{\bf A.} Frequency response of 3 conductance-based MSN model
\ neurons with variable scaling of IAs to uncorrelated input \ {\bf B.}
Probability distributions of ISIs. We see a clear separation of
frequency responses.}
\label{fig:fig1}
\end{figure*}

In Figure~\ref{fig:fig1}A, we show the response of individual, unconnected MSN model neurons 
with a scaling of
$\mu_{\mbox{\footnotesize IAs}} = 1.0, 1.3, 1.5$ to a noisy input signal, derived
from simulations of neural activity as uncorrelated Poisson-distributed spiking. 
The top panel shows the development of the membrane potential, 
$V_{m}$, over time for all neurons. 
The middle panel shows the spike-train for
each neuron with the mean interspike interval (ISI).
The bottom
panel shows the simulated synaptic input. The dots correspond to the spiking
events for a single neuron 3 ($\mu_{\mbox{\footnotesize IAs}} = 1.5$).
The resulting mean ISIs are 25, 37, and 45 ms. With a standard deviation
of 6, 11, and 8, they are clearly distinguishable. This is also shown by 
the Gaussian distribution for the mean ISIs for each neuron type 
(Figure~\ref{fig:fig1}B).

This model shows frequency-specificity as read-out of the relative
contribution of the slow A-type potassium channel, indicated by the
scaling factor $\mu_{\mbox{\footnotesize IAs}}$. The relative contribution of
an ion channel corresponds to its density or distribution on the
somato-dendritic membrane, or in some cases its specific localization
at dendritic branch points. Experimental evidence has shown that this
is a plastic feature for neurons.

\begin{figure*}[!h]
\showfig{
\centering
{\bf A}
\includegraphics[width=0.40\textwidth]{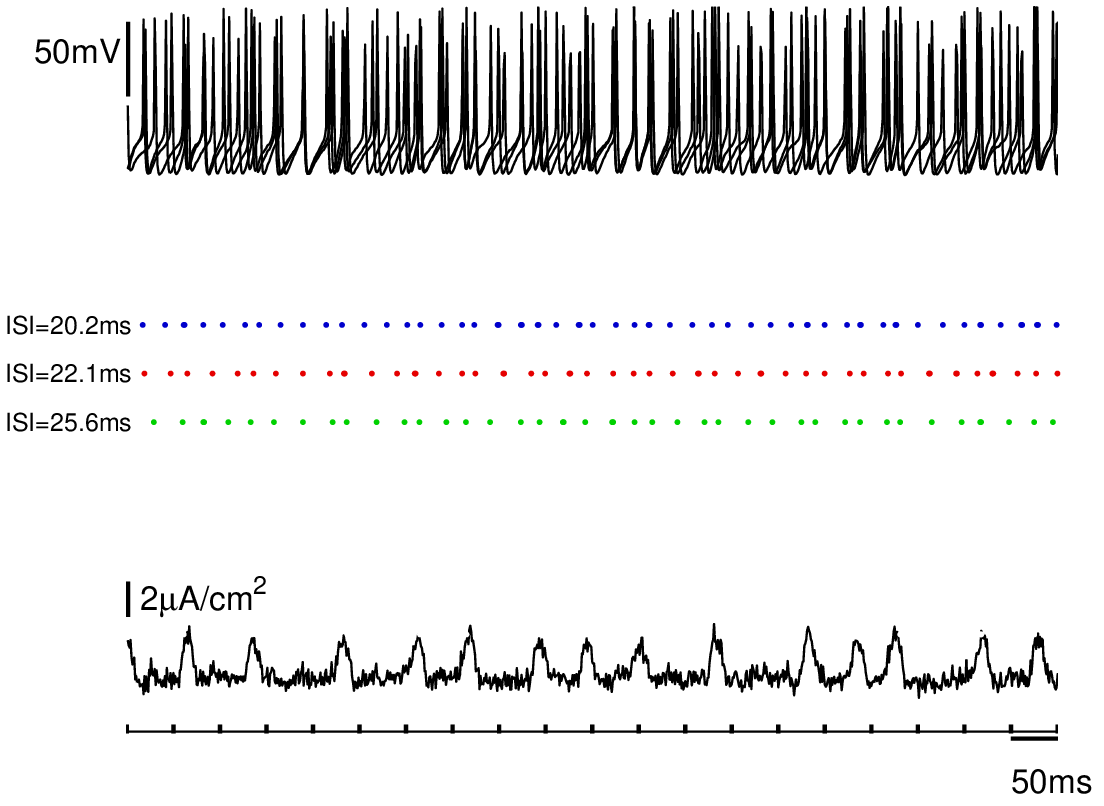}
\hspace*{1cm}
{\bf B}
\includegraphics[width=0.40\textwidth]{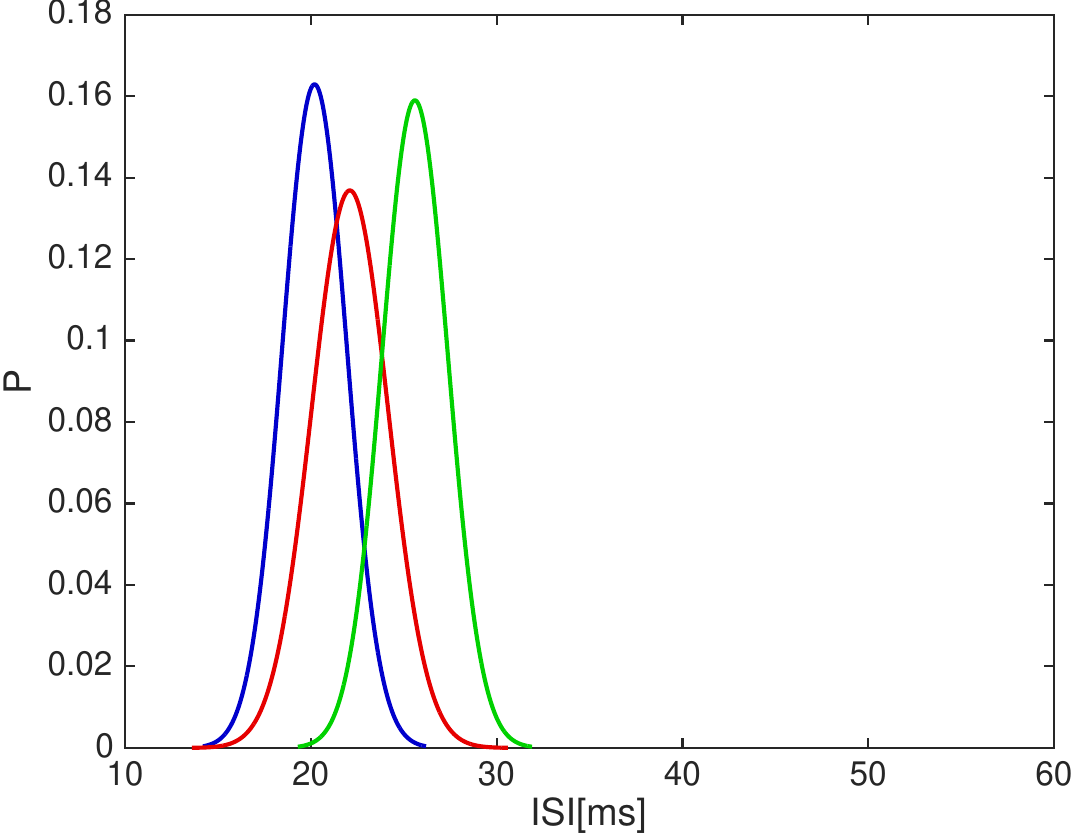}
}
\caption{
{\bf A.} Frequency response of the same MSN model neurons \ as in
Figure~\protect\ref{fig:fig1} to correlated input. 
{\bf B.} Probability distribution of ISIs. We see overlapping of frequency responses.}
\label{fig:fig2}
\end{figure*}

We then employ highly correlated synaptic input, defined as in 
\cite{Scheler2014} (see {\it Methods}). We stimulate the same neurons with the correlated input and 
observe the spike pattern (Figure~\ref{fig:fig2}). 
We can show that the frequency-specificity of the neuron disappears. 
Instead we see a time-locked spike pattern which
is expressed by a similar spike frequency (Figure~\ref{fig:fig2}A) and an 
overlap of the mean ISIs (Figure~\ref{fig:fig2}B).

What this experiment showed is that a stored intrinsic property, the
gain, is available to the processing network in a conditional manner.
The property is continually expressed, the differences in ion channel
density persist. Depending on the mode of stimulation, however, this
property is manifested as intrinsic gain, or it is obscured when a
neuron is driven by strongly correlated input.

\subsection{Results for simplified model neurons}
\begin{figure*}[!h]
\showfig{
\centering
{\bf A}
\includegraphics[width=0.40\textwidth]{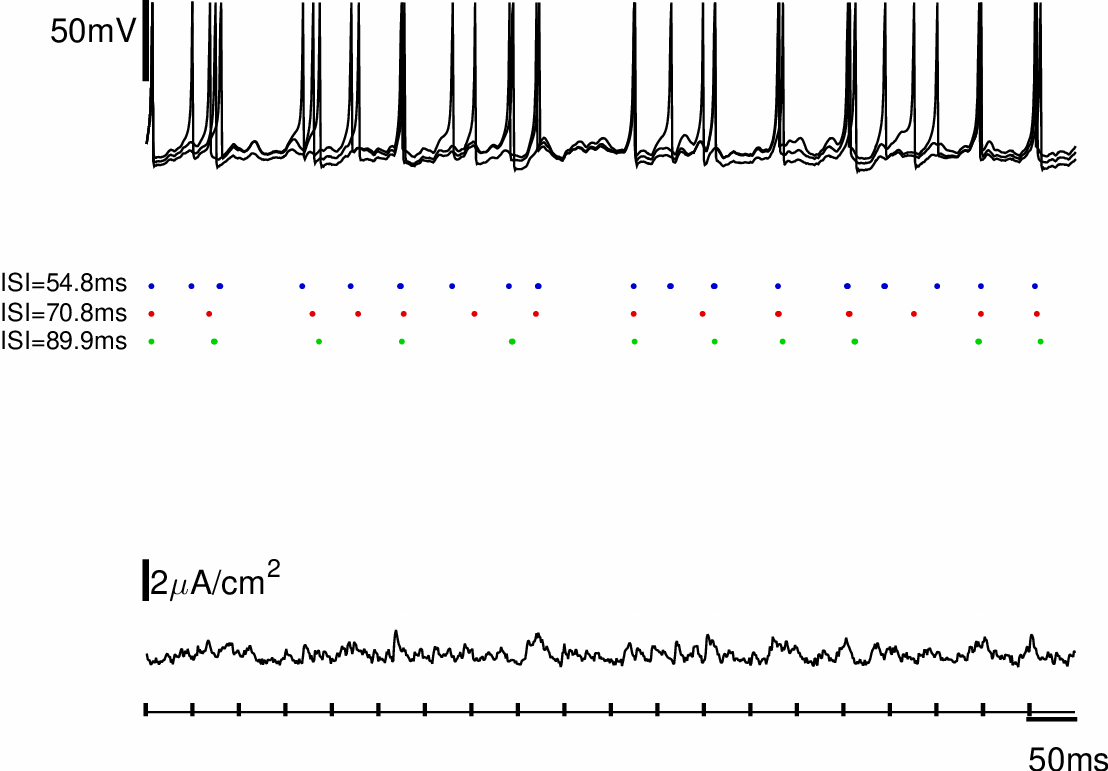}
\hspace*{1cm}
{\bf B}
\includegraphics[width=0.40\textwidth]{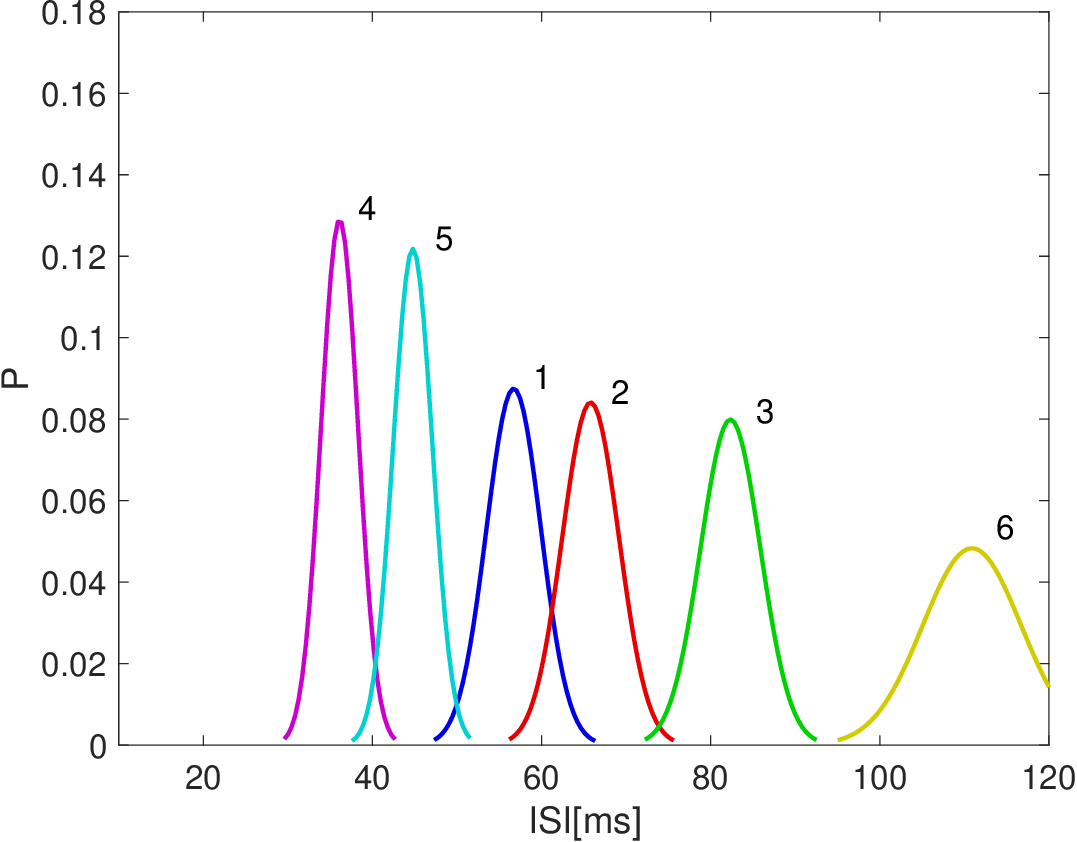}
}
\caption{
{\bf A.} Spike response of the two-dimensional dynamic model neurons
(1,2,3) to asynchronous input. 
{\bf B.} Gaussian distributions of ISIs for
six model neurons (1,2,3,4,5,6) to asynchronous input as in A. We see a
clear separation of frequency responses for model neurons.}
\label{fig:fig3}
\end{figure*}

\begin{figure*}[!h]
\showfig{
\centering
{\bf A}
\includegraphics[width=0.40\textwidth]{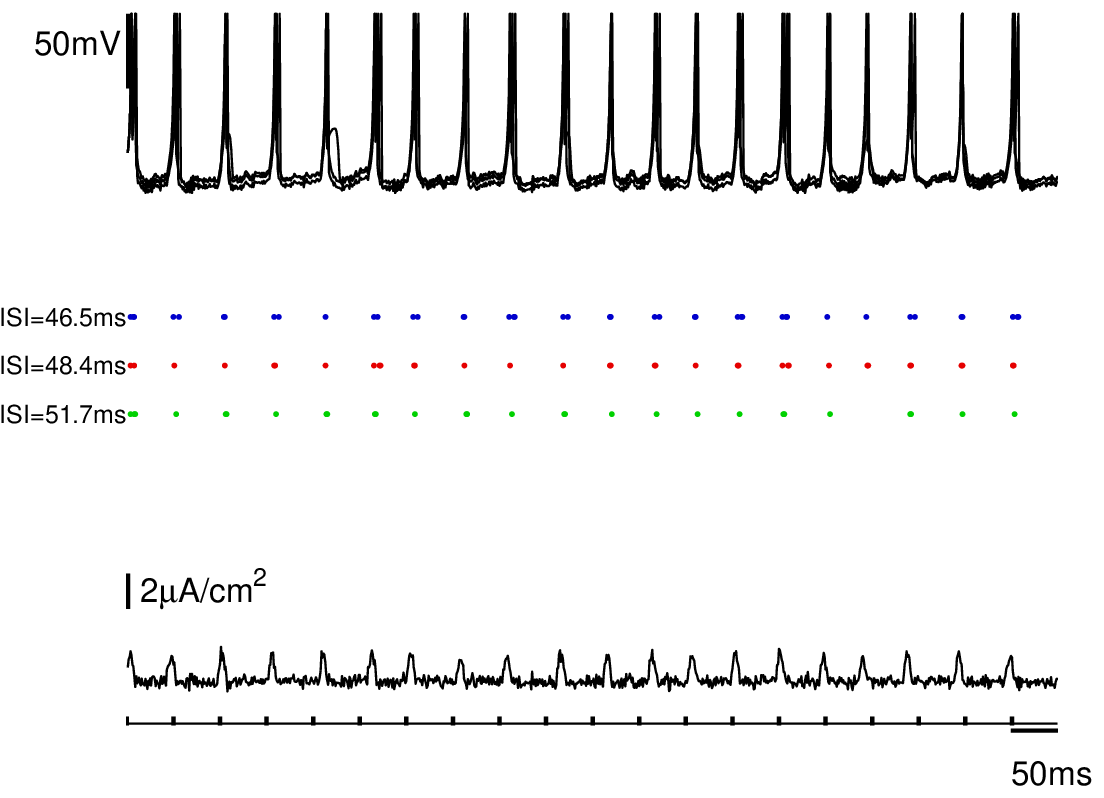}
\hspace*{1cm}
{\bf B}
\includegraphics[width=0.40\textwidth]{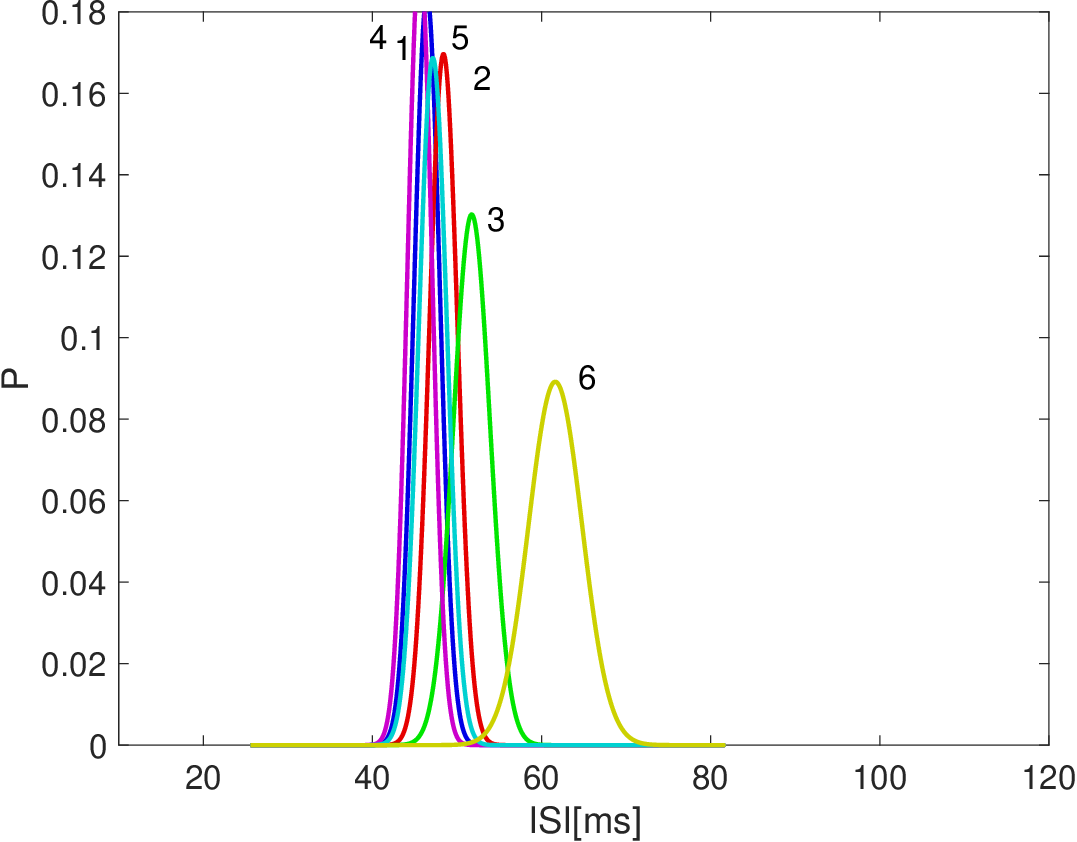}
}
\caption{
A. Spike response of the two-dimensional dynamic model neurons
(1,2,3) to regularly timed, correlated input. 
{\bf B.} Distributions of ISIs for six model neurons (1,2,3,4,5,6) to the
same input as in A. 
We see strong overlapping of frequency responses,
at about 50ms ISI, in accordance with the input. 
We notice that neuron 6 fires at lower frequencies than the input, 
it probably has a longer reset period, as seen in Figure~\protect\ref{fig:fig3}B.}
\label{fig:fig4}
\end{figure*}

\begin{figure*}[!h]
\showfig{
\centering
{\bf A}
\includegraphics[width=0.40\textwidth]{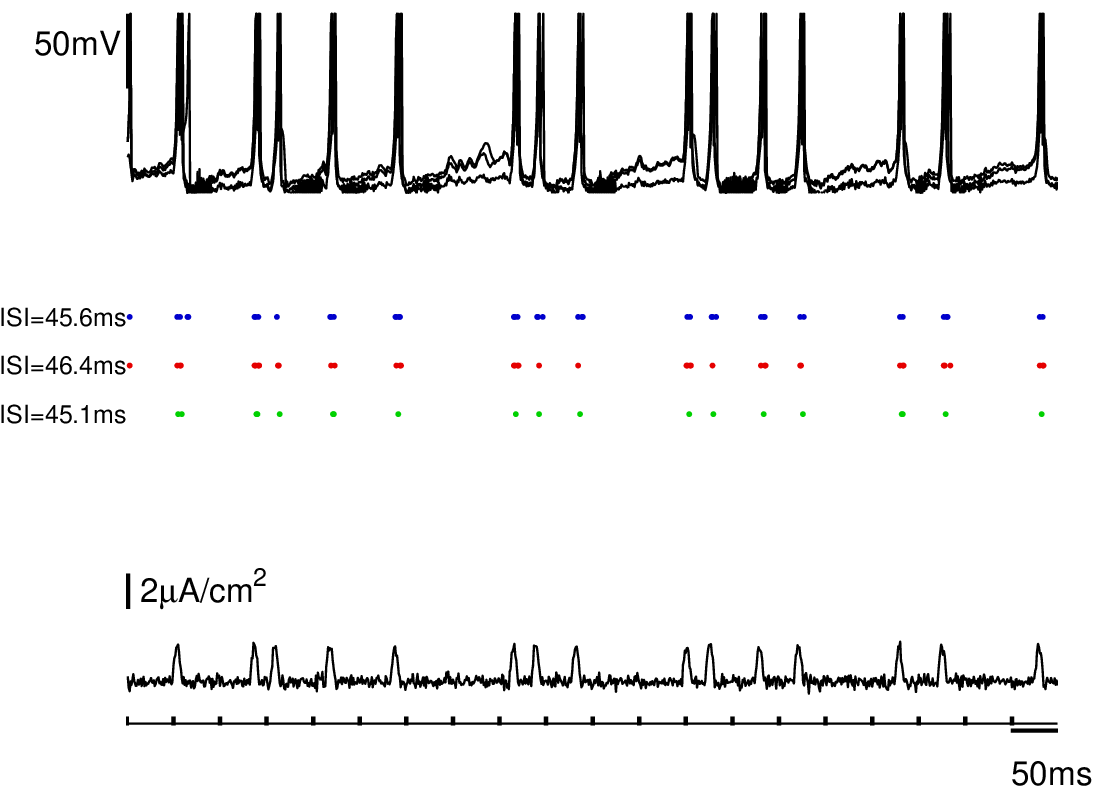}
\hspace*{1cm}
{\bf B}
\includegraphics[width=0.40\textwidth]{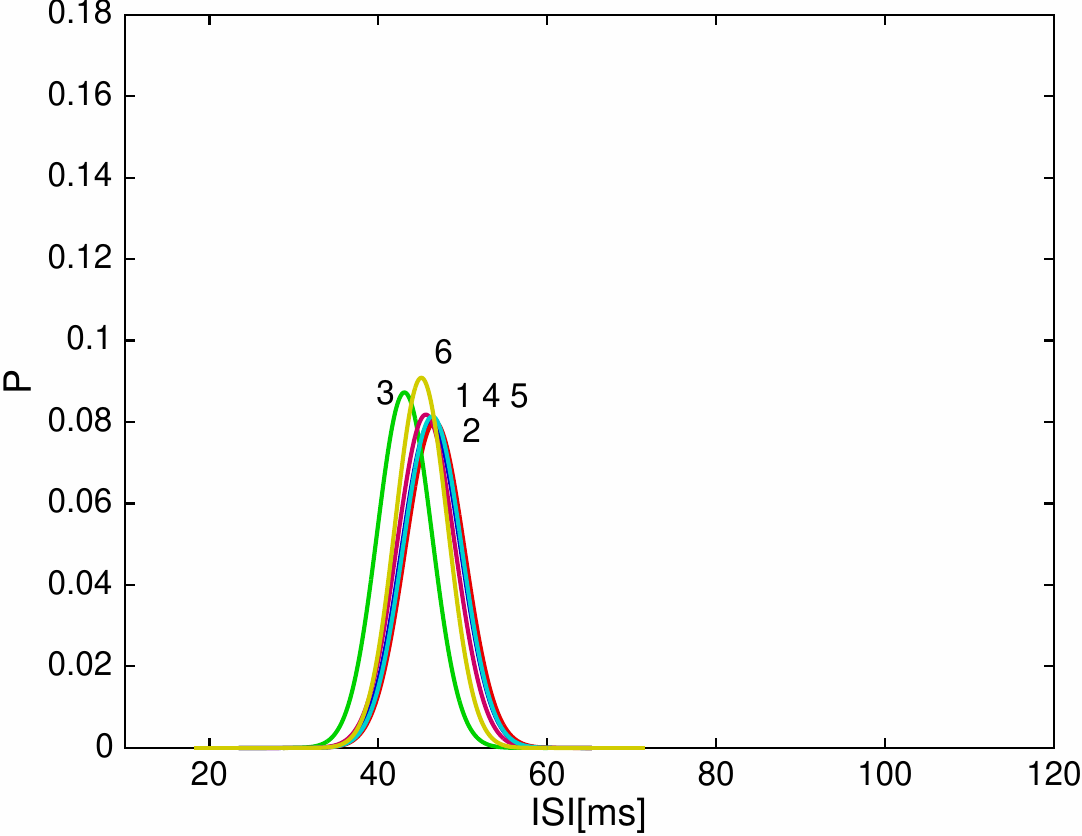}
}
\caption{
{\bf A.} Spike response of the two-dimensional dynamic model neurons (1,2,3) to 
irregularly timed, correlated input. 
{\bf B.} Gaussian distributions of ISIs for six model neurons (1,2,3,4,5,6). 
We see strong overlap of frequency responses.}
\label{fig:fig5}
\end{figure*}

To continue with exploring this property of model neurons, we switched to a simplified model neuron \cite{Izhikevich2004a} and created a set
of variations for this model (see \textit{Methods}). We
show the response of two-dimensional model neurons to asynchronous
input in Figure~\ref{fig:fig3}, and to regular, synchronous input in 
Figure~\ref{fig:fig4}. In the first case, we have clearly separated frequencies, 
and in the second
case, the ISIs are nearly identical with a narrow distribution. When we
stimulate the neurons with irregular, but synchronous input, the ISIs
become identical, but with a wider distribution to reflect the
different duration of pauses between the synchronous stimulation
(Figure~\ref{fig:fig5}).

\subsection{Multiplexing synchronous and asynchronous input}

\begin{figure*}[!h]
\showfig{
\centering
{\bf A}
\includegraphics[width=0.40\textwidth]{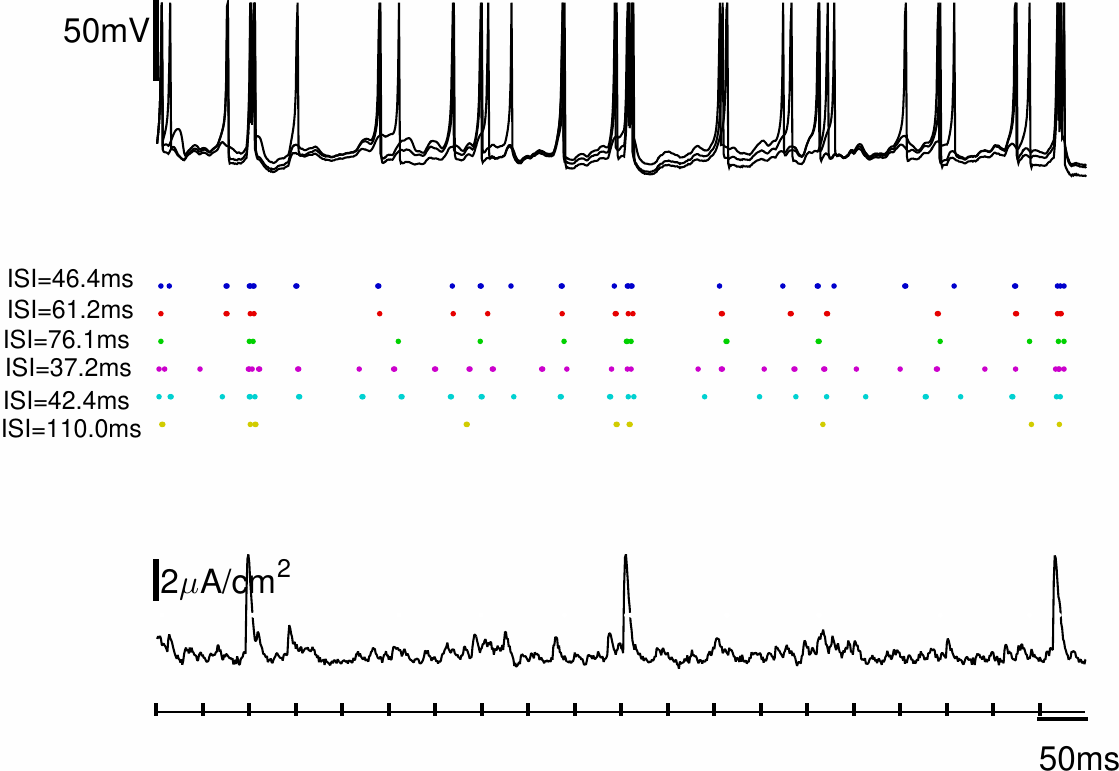}
\hspace*{1cm}
{\bf B}
\includegraphics[width=0.40\textwidth]{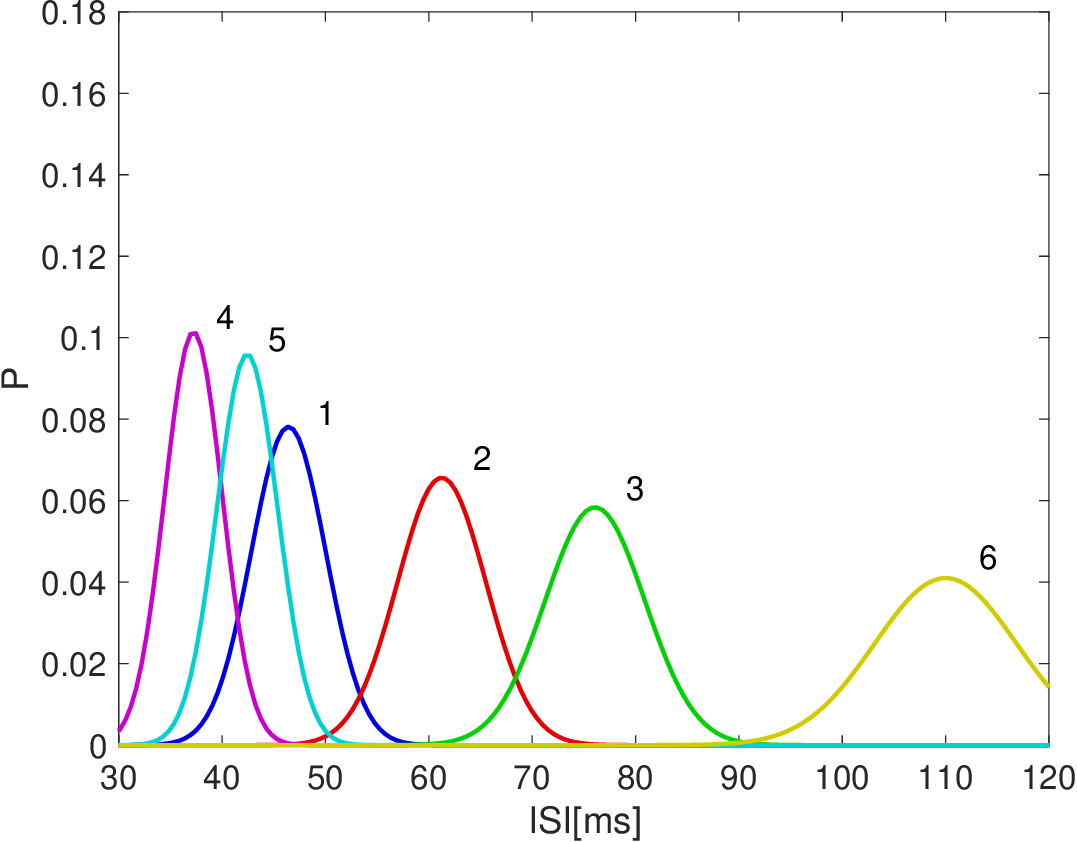}
}
\caption{
{\bf A.} Multiplexed input and response of different model neurons (1-6). 
Three synchronous events are clearly represented in  the spike pattern of 
all neurons. 
{\bf B.} Mean ISIs for each neuron type. The separation of frequencies is kept. 
Compared to Figure~\protect\ref{fig:fig3}B the standard deviation is somewhat 
higher because of the additional spikes caused by strong synchronous input.
}
\label{fig:fig6}
\end{figure*}

We may also consider the question of whether a neuron can simultaneously
respond to an input and read out its stored spike frequency. If there
are single synchronous events, which interrupt ongoing spiking,
can we recover the intrinsic properties for each neuron? 
In Figure~\ref{fig:fig6} it is
shown that this is possible. Figure~\ref{fig:fig6}A shows the input and the 
synchronous responses, and in Figure~\ref{fig:fig6}B we still 
see a clear separation of frequencies.

We conclude that we can multiplex asynchronous and synchronous input.  
It is also apparent that there needs to be a lower limit on the intervals 
between synchronous events that can be processed without disrupting intrinsic 
properties. 
This interval needs to be defined  as functionally dependent on the 
intrinsic frequencies. 
In this case, it is 3/s for the synchronous events, with 10Hz for the 
slowest neuron.

\subsection*{Synchronization depends on network topology}

The simplified model neurons allow the creation of large networks of
heterogeneous neurons and exploration of different topologies (cf. also \cite{Hu2014, Trousdale2012}). We hypothesized
that a lognormal graph, because of its hierarchical topology and the
existence of hub neurons would lead to synchronization of action
potentials -- even with heterogeneous neurons -- while a Gaussian
topology would support asynchronous spiking behavior \cite{Barrat2008,Arenas2008}. We define
synchronization $s$ in a network by pairwise correlation
(\textit{Methods}). The spike frequency for each
neuron type is assessed by the mean and standard deviation for ISIs, as
before.

\begin{figure*}[!h]
\showfig{
\centering
{\bf A}
\includegraphics[width=0.30\textwidth]{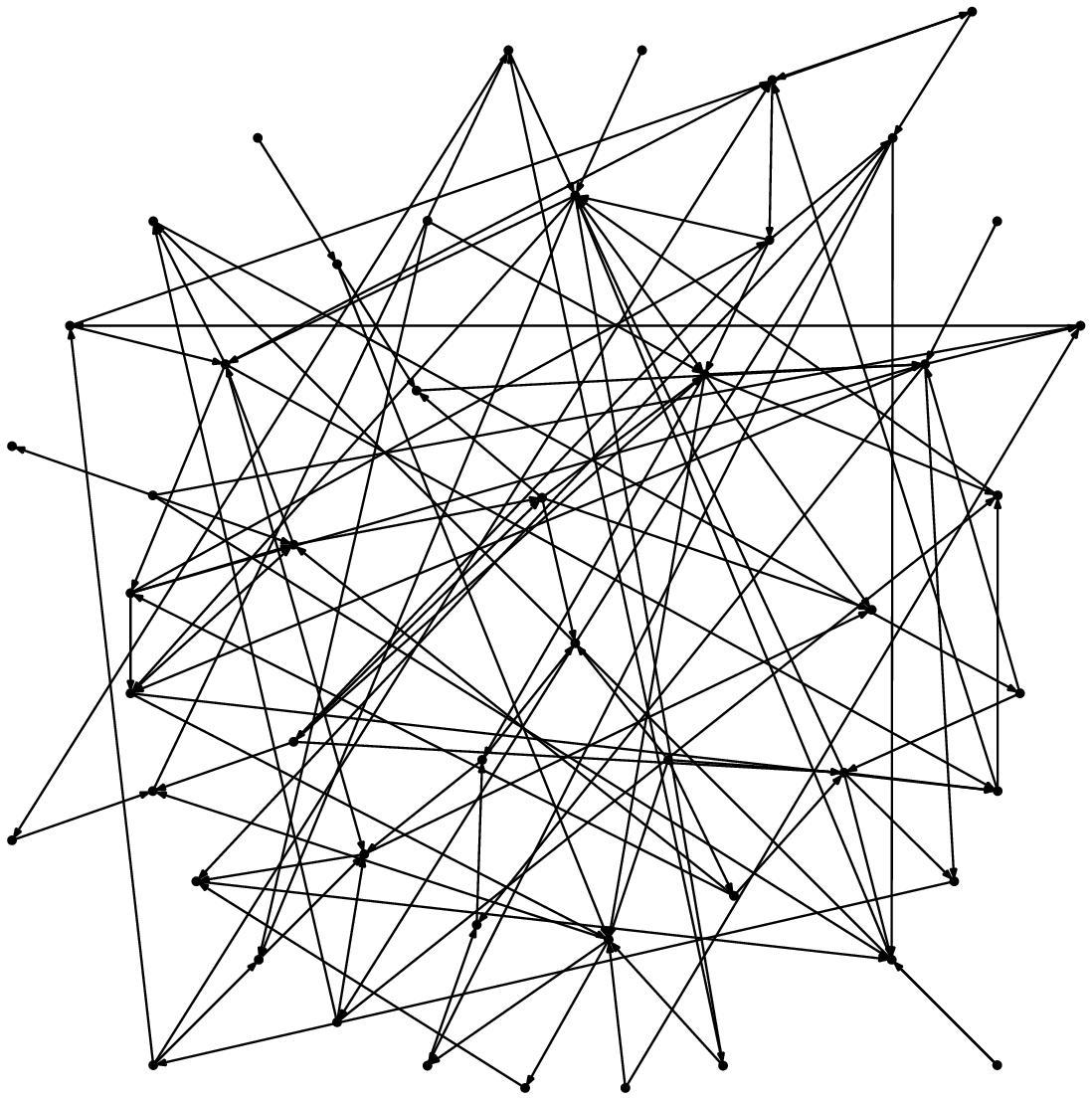}
\hspace*{1cm}
{\bf B}
\includegraphics[width=0.30\textwidth]{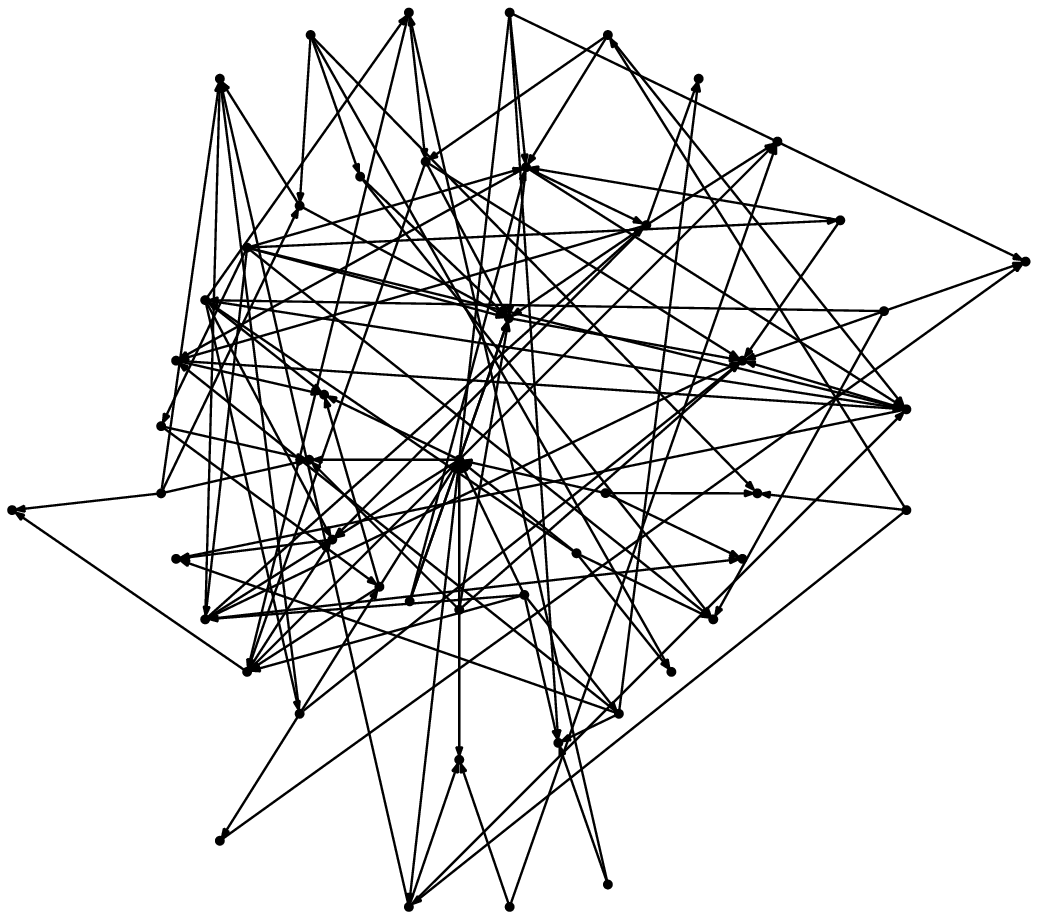}
}
\caption{
{\bf A.} Part of a Gaussian Graph (RG), here for 50 neurons  
{\bf B.} Part of a Lognormal Graph (LG1), for 50 neurons.  The more regular, lattice-like structure of the Gaussian graph and the higher clustering and the appearance of highly connected 'hub' neurons in the lognormal graph is apparent. 
}
\label{fig:fig7}
\end{figure*}

We first use a randomly (Gaussian) connected graph (RG) with 210 neurons ($N=210$) and 1800 excitatory connections ($K=1800$). We employ 7 different neuronal types (1-6, plus the generic neuron $g$) with 30 Neurons each (\textit{Methods}).
Figure~\ref{fig:fig7}A shows an excerpt of the graph structure. 
We can see that the
graph is connected such that all neurons have a comparable number of
connections. This is also apparent in Figure 8, where we can see a
(narrow) normal distribution for connectivity for the Gaussian graph
RG. Table~\ref{tab2} contains the usual graph characteristics.

$N=210$ is about the size of a minicolumn or ensemble unit within a larger
network with presumably dense interconnections \cite{Klinshov2014}. 
The maximal density $d=K/(N\times (N-1))$ in a cortical microcircuit 
is estimated at 0.1 for $10^4$ neurons, and $10^7$ synaptic connections, \cite{Lansner2012}.
With $\approx 50\%$ of synapses internal to the network, $d=0.04-0.07$ is a
realistic value for internal connectivity \cite{Klinshov2014}. There is also a small background inhibition to
all neurons present, implemented by 10\% inhibitory neurons with
Poisson-distributed firing and complete connectivity to excitatory
neurons.

\begin{figure*}[!h]
\showfig{
\centering
\includegraphics[width=0.98\textwidth]{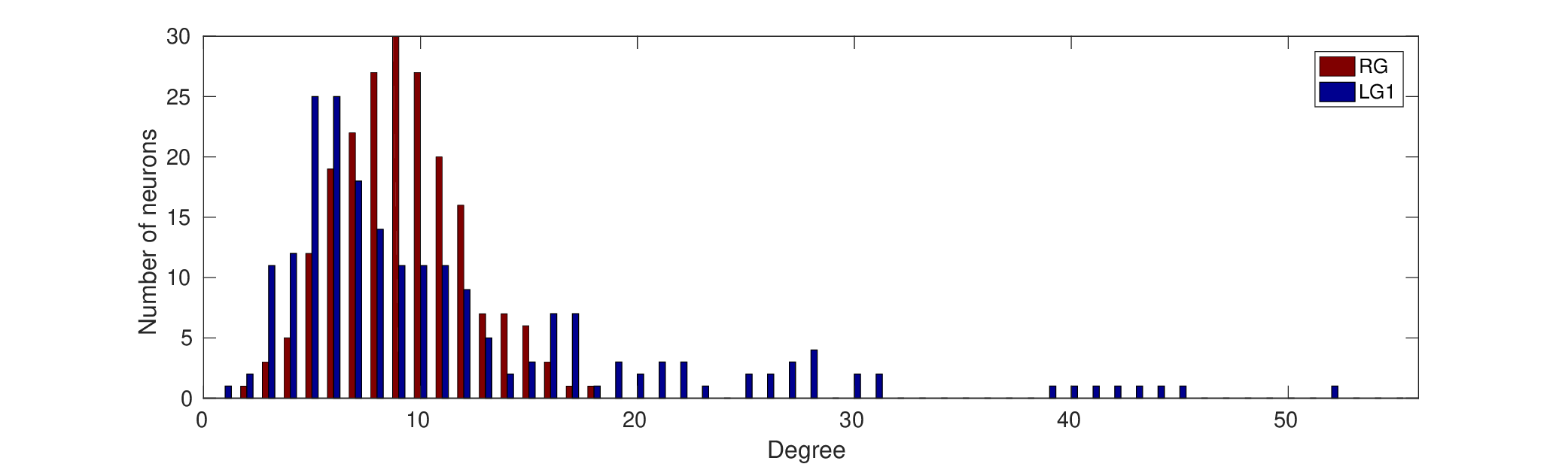}
}
\caption{
Degree histogram for the Gaussian Graph RG (red) and the lognormal Graph LG1 (blue).  The LG has more neurons with few connections. It also has a heavy tail of neurons with 20 and more connections ('hubs'), which are lacking in the Gaussian graph.
}
\label{fig:fig8}
\end{figure*}

\begin{figure*}[!h]
\showfig{
\centering
{\bf A}
\includegraphics[width=0.40\textwidth]{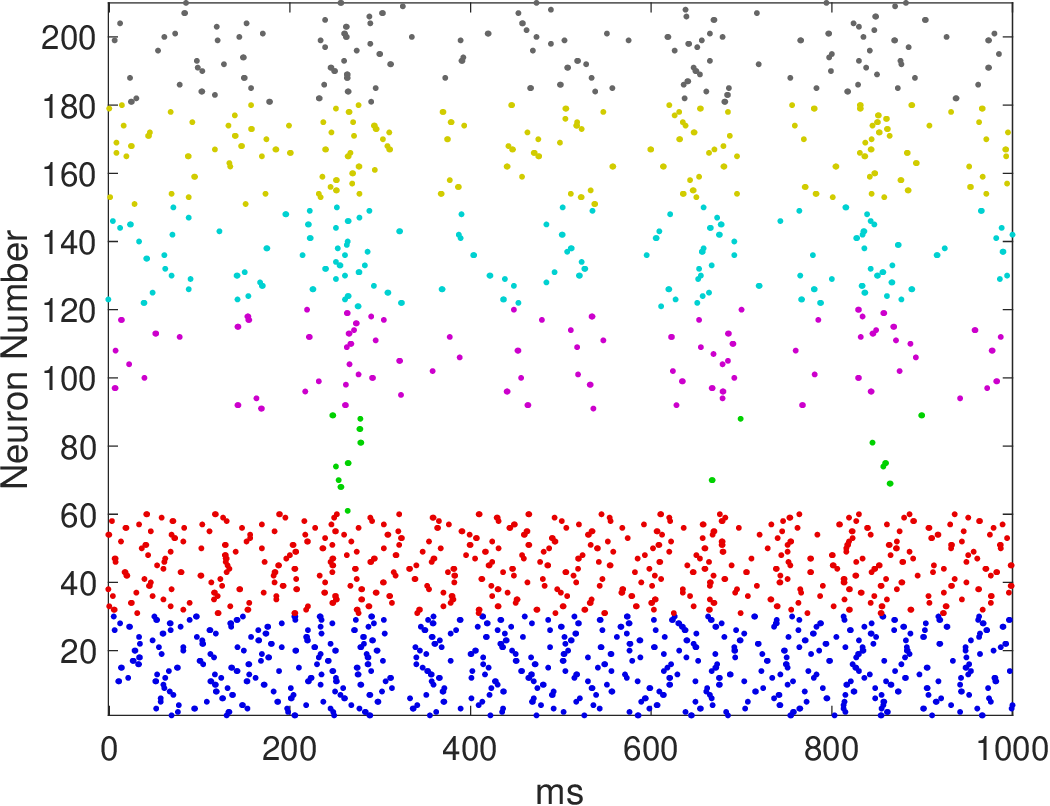}
\hspace*{1cm}
{\bf B}
\includegraphics[width=0.40\textwidth]{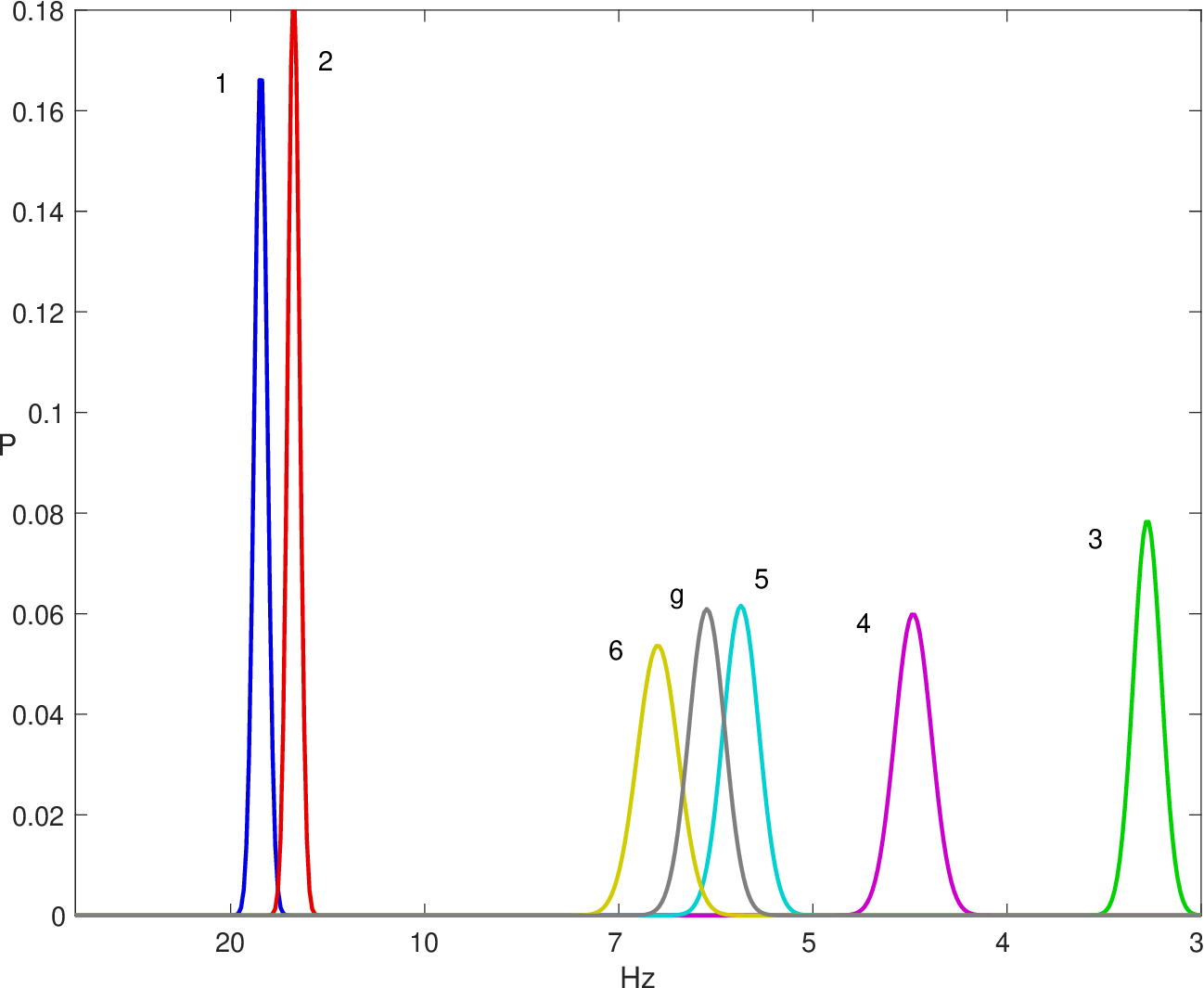}
}
\caption{
{\bf A.} Asynchronous behavior in the Gaussian graph (RG) with variable
neuron types. 
Groups of neuronal types with 30 neurons each are apparent in the rasterplot. 
Some structure is probably due to background inhibition. 
{\bf B.} Average spikes/s with probability distributions for all neuronal
types, with clear separation by frequency. 
Pairwise correlation is $s = 0.11$.}
\label{fig:fig9}
\end{figure*}

We now stimulate the graph by an initial stimulation to 10 randomly selected excitatory
neurons (for about 1 second).
In Figure~\ref{fig:fig9}A, we see highly asynchronous neuronal activity
after 1s of stimulation. The pairwise correlation value $s$ is low
($s=0.11$). 
Figure~\ref{fig:fig9}B shows that each neuronal type retains its own
frequency, i.e., has its own typical ISI, separated from other neuronal
types. We also notice that some neurons fire with low frequencies (5Hz)
and others with higher frequencies (20Hz).
Very low firing neurons (2Hz) which are typical for cortex are not 
represented in this model.

Next we changed the topology of the network to a graph with a lognormal distribution of 
connections (LG1), as shown in Figure~\ref{fig:fig8}.
It used the same neurons ($N=210$) and approximately the same number of excitatory connections $K=1924$ as before.

Figure~\ref{fig:fig7}B shows an excerpt of the lognormal graph structure from LG1. The
connectivity structure seems much denser, because of
'hub' neurons in the center of the graph. 
In Figure~\ref{fig:fig8}, we can see the wider distribution of degrees for
the lognormal graph (blue), containing a number of nodes with high
connectivity ('heavy-tailed distribution'). Presumably, those nodes are 
capable of synchronizing the
network, because they can reach many neurons simultaneously. What is
the effect on the presence of neural heterogeneity?

\begin{figure*}[!h]
\showfig{
\centering
{\bf A}
\includegraphics[width=0.45\textwidth]{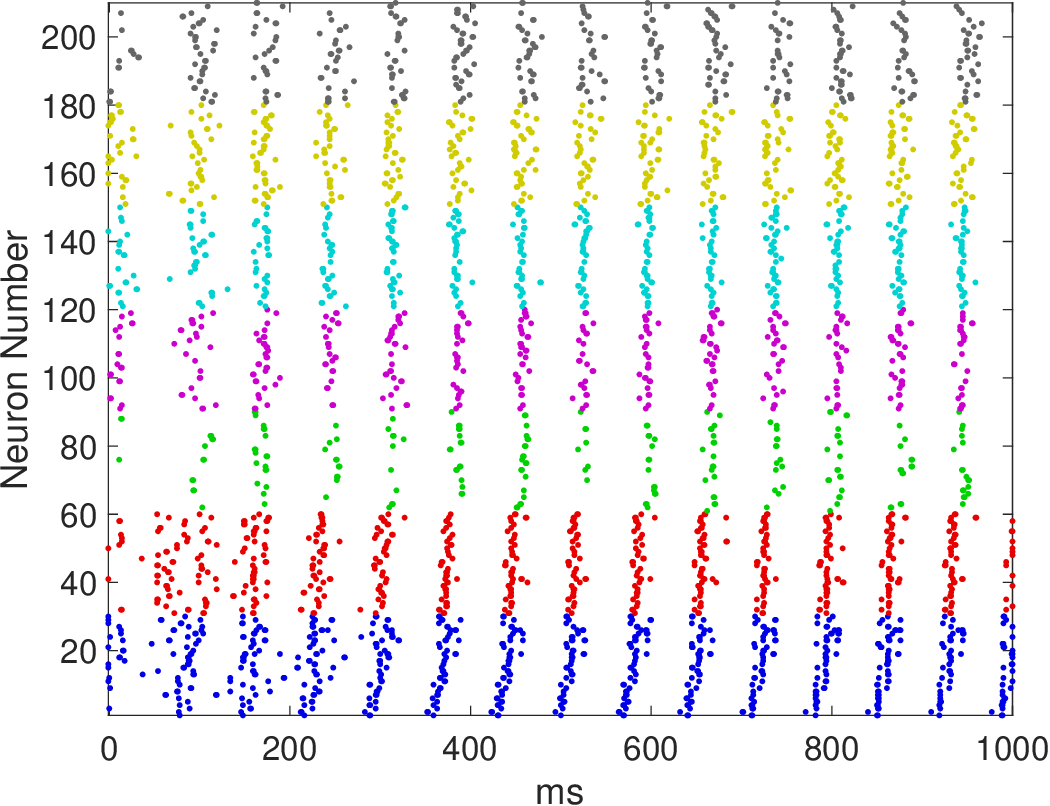}
\hspace*{1cm}
{\bf B}
\includegraphics[width=0.40\textwidth]{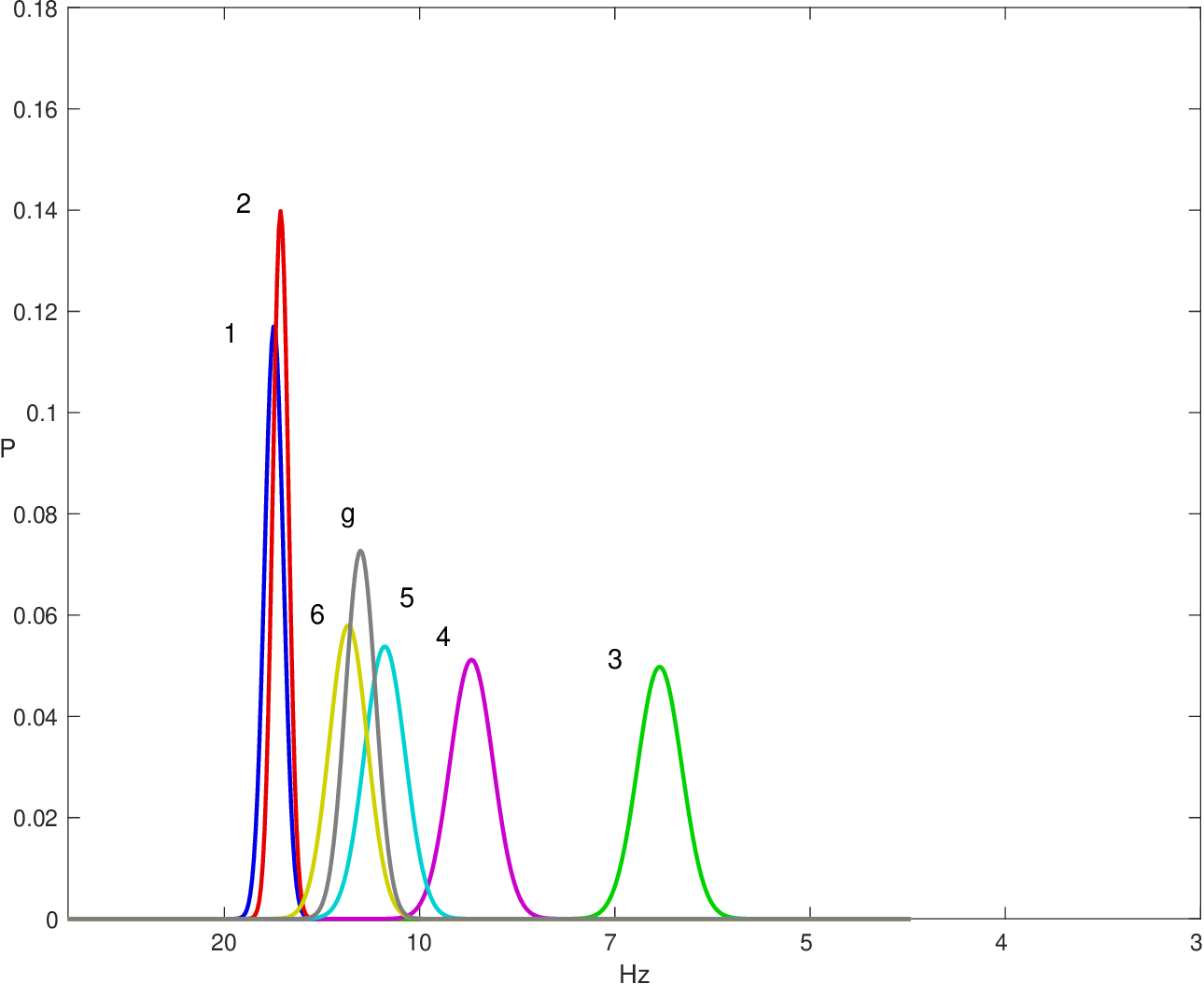}
}
\caption{
{\bf A.} Synchronization in a heavy-tail graph (LG1) with variable
neuron types. The rasterplot shows that different neuronal types
respond uniformly. 
{\bf B.} Frequency distributions. High overlap between
neuronal types is apparent. Pairwise correlation in the graph is high
with $s = 0.32$.}
\label{fig:fig10}
\end{figure*}

Figure~\ref{fig:fig10} shows that a high amount of synchronization can be achieved in spite of heterogeneity of intrinsic frequency of model neurons. 
The rasterplot
(Figure~\ref{fig:fig10}A) shows the activity in LG1 with the same neurons 
and the same stimulation as before. The overall correlation, defined by 
pairwise correlation of neurons, is much higher (s=0.32). The distribution of
ISIs in this case is strongly overlapping (Figure~\ref{fig:fig10}B), 
similar to Figure~\ref{fig:fig4}B, where neurons were explicitly driven 
by highly synchronous input.
This means that synchronization is dependent on the network topology,
and a lognormal graph exhibits a higher tendency for pairwise
synchronization. Also, that neuronal heterogeneity is apparent in an
asynchronous network mode but is repressed in a
synchronous firing mode.

\subsection{Dependence of synchronization on graph properties}

We could show that differences in intrinsic properties appear or 
become more prominent when there is less synchronicity in a network. 
In our model, the pairwise synchronicity $s$ is dominated by the network
topology, more precisely by the width of the degree distribution
ranging from Gaussian to lognormal with a heavy tail.

\begin{figure*}[!h]
\showfig{
\centering
\includegraphics[width=0.79\textwidth]{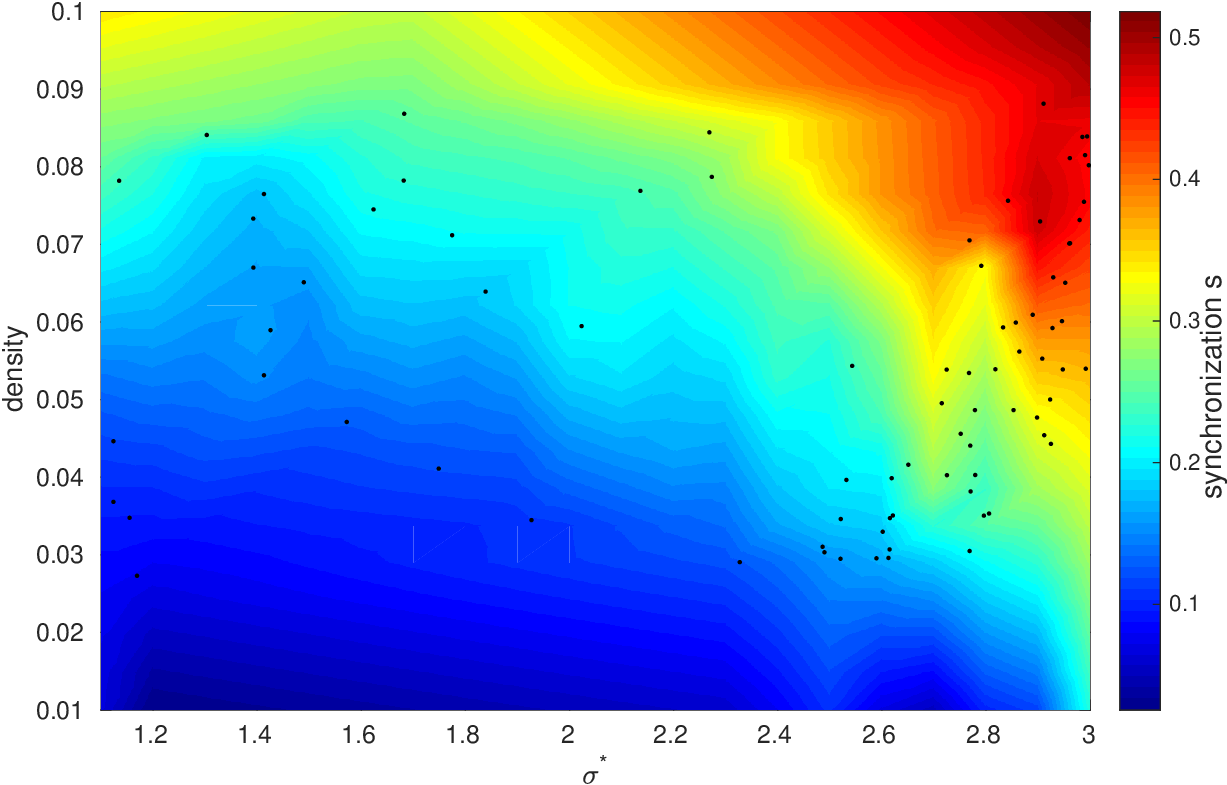}
}
\caption{
Synchronization $s$ dependent on network topology: density and distribution
width $\sigma^*$. 
The experimentally attested width for weights in cortical tissue
\cite{Scheler2017} is $\sigma^*= 2-3.5$, with a mean at 3. 
We achieve higher synchronization $s$ in the lognormal region, also dependent 
on density, but no synchronization in the Gaussian (region of low width)
($\sigma^*<1.5$), except close to maximal connectivity. 
Black dots signify actual measurements. There seem to be no abrupt transitions.
}
\label{fig:fig11}
\end{figure*}

To confirm this observation we used a number of intermediate graphs 
and mapped the pairwise synchronization dependent on the distribution width 
$\sigma^*$ (Figure~\ref{fig:fig11}).
The graphs RG and LG1 that we used have values of $\sigma^*=1.44$
 and $\sigma^*=2.89$ ({\it Methods}). 
They have the same density, i.e., the same number of connections and 
neurons ($\approx 0.05$).  
Additionally, we analyzed the dependence of synchronicity on the density of 
the graph between 0.01 and 0.1 (Figure~\ref{fig:fig11}).

There is higher synchronization in the lognormal region, especially with 
higher $\sigma^* > 2.5$  but no synchronization for Gaussian graphs. 
For heavy-tail graphs, synchronization depends linearly on the density 
between $d=0.03-0.08$ ($s=0.2-0.5$).

\begin{figure*}[!h]
\showfig{
\centering
\includegraphics[width=0.79\textwidth]{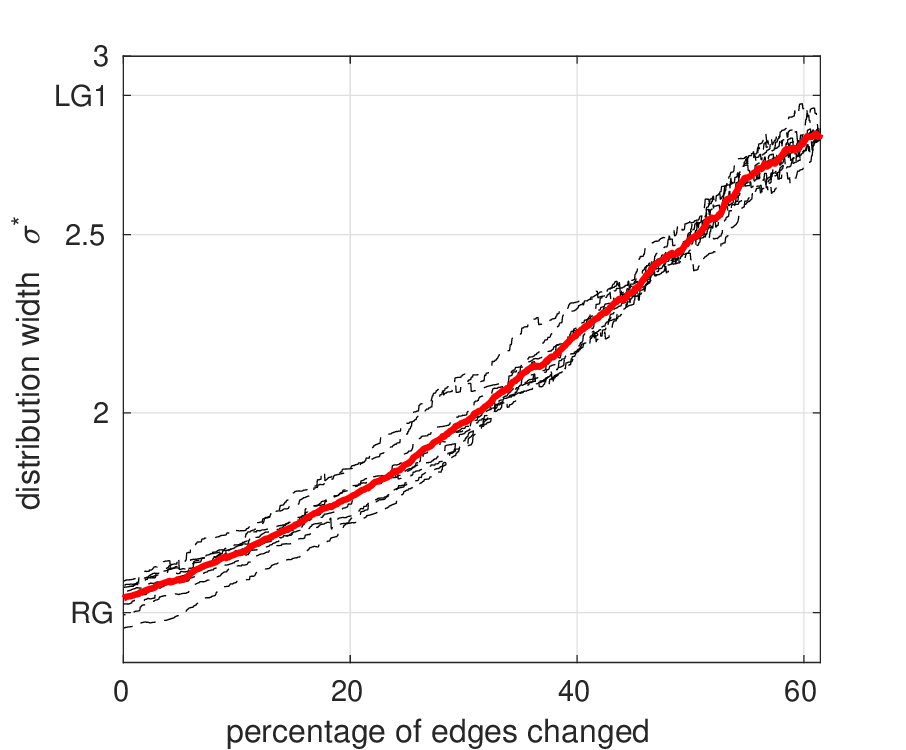}
}
\caption{
Transition between a lognormal graph and a Gaussian graph: For K=1900, N=210, d=0.43, mean over 10 trials, the percentage of edges changed to achieve a narrow degree distribution. The algorithm is not optimized (Methods) and overestimates the number of edges that have to be changed.
}
\label{fig:fig12}
\end{figure*}

How are the different graphs related? We hypothesized that fast synaptic
switching \cite{Scheler2003d} by neuromodulation could change the network topology
sufficiently to switch from a synchronous to an asynchronous regime. 
In Figure~\ref{fig:fig12}, we plot the number of edges that were changed 
to achieve different distribution width $\sigma^*$ of a graph. 
The algorithm used was a simple greedy algorithm 
(\textit{Methods}), which is
suboptimal, i.e., overestimates the number of edges required. It appears
that 30--50\% of edges changed would be sufficient.

\section{Discussion}

\subsection{Network Topology, Synchronization and Intrinsic Read-out}

We employ a parameterizable two-dimensional neural oscillator model to
encode different intrinsic excitability manifested by different
frequency responses to constant input.
What the experiments show is that a stored intrinsic property, the
gain, is available to the processing network in a conditional manner:
the gain is continually present, the differences in ion channel density
persist. Depending on the mode of stimulation, however, this property
is manifested as spike rate, or it is obscured when a neuron
is driven by strongly correlated input. This is interesting because it
shows a property of memory that synaptic plasticity lacks: the memory
is not always 'read-out' in any
processing step. It is conditional, it can be accessed or ignored
depending on the state of the network. This seems to be an essential
property of memory in any intelligent system.

Different statistical properties of synaptic input can be modeled by a
variability in the correlation properties of input neurons. In a
network model, this means that the overall correlation in the network
determines what input a neuron receives. With a Gaussian degree
distribution topology, correlation is low and neurons fire irregularly
with their own preferred frequency. With a heavy-tailed, lognormal degree distribution
topology, correlation is higher, and neurons fire when they receive
correlated input, irrespective of intrinsic properties. I.e., driving
neurons by correlated vs.~uncorrelated input leads to uniform spiking
behavior vs. read-out of stored differences in ion channel
conductances.

\subsection{Inhibition}

A restriction of the present model with respect to a biological
simulation model is the simplified treatment of inhibition. However,
\ experimental work shows that cortical parvalbumin-expressing (PV+),
fast-spiking interneurons have no connection specificity to pyramidal
neurons, rather they present as an 'unspecific,
densely homogeneous matrix covering all nearby pyramidal
cells' (\cite{Packer2011}, p.~13260), 
which corresponds to our model.

Conditions for neuronal read-out may include the activity of inhibitory
neurons. Inhibition and excitation are tightly linked by feedback
interaction. \cite{Graupner2013}
suggested that the close coupling of inhibition and excitation in cortical tissue cancels out purely input-dependent, i.e. not network generated synchrony.
\cite{Rudolph2003} suggested that with highly correlated input, both inhibitory and
excitatory, the neuron may receive less input which allows it to be driven only by strong synaptic input, while distributed input consists of a barrage of excitatory and inhibitory inputs where
the membrane voltage remains close to firing threshold and the neuron
fires continuously. In our sense, it is 'reading out' its stored intrinsic frequency. 
Inhibitory and excitatory synaptic input conform to be either asynchronous or synchronous, to drive neurons by correlated input or to 
cause them to emit spikes according to their own intrinsic frequency.

However, neuromodulation has effects on inhibitory neurons as well
\cite{Kruglikov2008,Stringer2016}, which we have not modeled. 
Further simulations will show whether the I-E coupling is altered during
enhanced neuromodulation, or whether the effects are synergistic with the
present results.

\subsection{The role of neuromodulation}

Neuromodulation influences both intrinsic properties and synaptic
connectivity \cite{Scheler2004g}, e.g., acetylcholine, (via nucleus basalis
stimulation), noradrenaline (via LC stimulation) or dopamine (via VTA
stimulation) \cite{Guedj2016,Scheler2004g}. Experimental estimates on the
distribution of synaptic neuromodulatory receptors are at approximately
30\%--50\% of connections \cite{Scheler2003d}. That is sufficient to transform the
topological properties of a graph, such as the width of its degree
distribution from heavy-tailed graph to a more Gaussian, less
clustered graph without requiring tight optimization for the positions
of neuromodulatory receptors (Figure~\ref{fig:fig12}).

Neuromodulation disables or enhances various ion channels, such as
Sk-channels which guide reset times after a spike, or A-type
potassium channels which influence latency to spike \cite{Vogalis2003,Nadim2014,Scheler2014}.
In this way,
neuromodulation influences intrinsic properties \cite{Nicoll1988}. If neuromodulation
reduces synchrony by acting at synaptic receptors, it uncovers
intrinsic heterogeneity, and induces a mode of processing that allows
read-out and storing of intrinsic properties. Depending on the
neuromodulator used, and the amplitude and duration of the signal, different soma-dendritic ion channel profiles
would emerge \cite{Khorkova2013,Li2018}.

In the synchronous mode, intrinsic heterogeneities are reduced in the
presence of tightly correlated input which drives neurons reliably.
This invariance of neuronal intrinsic properties in synchronous mode
allows synaptic transmission and information processing independent of
neuronal heterogeneity.

The idea of introducing synchronous events by common input to an
asynchronous background, and in this way use reliable synaptic
transmission without affecting the state of the system (multiplexing)
has also been documented in experimental results. For instance,
(\cite{Gutnisky2017}, Figure~4A)
shows a case of multiplexing in response to
behavioral stimuli. In this case, intrinsic read-out can continue, and
single events are transmitted reliably through driven activations.

Why should synchronization properties be switched by neuromodulation?
Increased correlation in the network supports population-coded information to be propagated effectively \cite{Zylberberg2017}.  Turning on neuromodulation would decorrelate an area and increase the capacity for information coding in an ensemble or a cortical microcolumn \cite{Sompolinsky2001}. This area would become an information source to surrounding areas. When turned off, increased correlation would allow this area to transmit information and to disregard the stored neural memory.

\subsection{Relation to experimental evidence}

Basal forebrain stimulation, which results
in increased acetylcholine release and muscarinic/nicotinic receptor
activation, decreases correlation between cortical neurons
(\cite{Goard2009,Edeline2012}, \cite{Lee2012b} (Figure~3.C)). 
Likewise,
\cite{Minces2017a} (Figures~3 and 4) 
shows reduced noise (internal)
correlations with cholinergic stimulation, while inactivation of the
basal forebrain caused more synchronized activity. 
\cite{Jeanne2013} shows
reduction of correlation for task-relevant perception, where presumably
task-relevance causes neuromodulatory activity. 
\cite{Fazlali2016} provides evidence for the involvement of
noradrenaline in
desynchronization of cortical state and the enhancement of sensory
coding.

There is considerable evidence 
\cite{Harris2011,Scholvinck2015,Renart2010,Beaman2017} 
showing that several neuromodulators,
including at least noradrenaline and acetylcholine, modulate pairwise
spike correlation, such that strongly synchronized states (anesthesia,
slow wave sleep) have high correlation and low neuromodulation, while
asynchronous states (normal waking), with higher neuromodulation, have lower pairwise correlation. 

\cite{Beaman2017} observed intrinsic
fluctuations in synchronization of cortical networks during wakefulness
{which correlated with the amount of
encoded perceptual information and perceptual
performance.
Their results showed
a mean decrease in correlations from synchronized to desynchronized state corresponding to perceptual performance
by approximately 20\%, similar to
values observed during attention
\cite{Mitchell2009},
and after adaptation \cite{Gutnisky2008}.
We have shown (Figure~\ref{fig:fig11}) that correlation changes are continuous 
with network topology and a 20\% correlation change is well within the range
of the current simulations.
Importantly, the results in
\cite{Beaman2017}
point to fluctuations in synchronization that reflect
local changes in network activity rather than just global cortical state
dynamics which have traditionally been associated with
central neuromodulatory release.

The role of presynaptic neuromodulation in suppressing
cortical connections
\cite{Ohshima2017,Kobayashi2009} and changing attractor states \cite{Kanamaru2013}, as well as
allowing rapid synaptic weight changes \cite{Scheler2003d} has previously been
assessed. Theoretical work has also emphasized the connection between
correlations and information content 
\cite{Cohen2011,Sompolinsky2001,Cohen2008,Poulet2008}.

Here we bring these observations together to suggest that neuromodulation of synapses may alter network topology and in this way bring about an increased decorrelation of spiking, and a more asynchronous
state, with a higher informational capacity.
It may provide a general explanation (a) on how
fluctuations in synchrony can be engineered rapidly and in small
cortical areas and (b) why intrinsic memory may be conditional,
accessible only at certain times and in a localized fashion.

\section{Conclusion}
We created a number of different parameterized neuron models to capture
neuronal heterogeneity. This affects the properties of the neuron such
that it has less or more intrinsic excitability, leading to different
firing rates when stimulated in an asynchronous way. Under synchronous
stimulation the differences are greatly reduced.

We also suggested that synaptic neuromodulation can be an effective way
of rapidly altering network topology. We investigated changes in
network topology along the dimensions of Gaussian vs.\ heavy-tailed
degree distributions. We hypothesized that heavy-tailed graphs produce
more globally synchronized behavior than comparable Gaussian graphs.
In accordance with the hypothesis, we find that in a heavy-tailed
graph, because of high population synchrony, the difference between
neuronal intrinsic properties is minimized, while a Gaussian graph
allows read-out of neuronal intrinsic properties. Thus, altering
network topology can alter the balance between intrinsically determined
vs. synaptically driven network activity.

\comment{
\subsection*{Data availability}
Underlying data for this study is available from Zenodo. Dataset 1. gscheler/CNeuroSyn: initial version, \url{https://doi.org/10.5281/zenodo.1164096}.
%
%

Data is available under a Creative Commons CC BY-NC 4.0 license. 
%
%

\subsection{Software availability}
The source code for the model is available from GitHub: \url{https://github.com/gscheler/CNeuroSyn/tree/V1.0/src/analysis} \\

Archived source code at time of publication is available from Zenodo \url{https://doi.org/10.5281/zenodo.1164096}. \\

Software is available under GNU GPL v2.0 license.
%
%

\subsection*{Competing interests}
No competing interests were disclosed

\subsection*{Grant information}
The author declared that no grants were involved in supporting this work.
}

\clearpage

\newcommand{\commenturl}[1]{}

{\bibliographystyle{unsrt}


}

\end{document}